Stanford | **Stanford Existential Risks Initiative**
Center for International Security and Cooperation
*Freeman Spogli Institute*

# Fairness in AI and Its Long-Term Implications on Society

Ondrej Bohdal, [1*] Timothy Hospedales, [2] Philip H.S. Torr, [3] Fazl Barez [4]

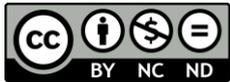

**Funding:** Ondrej Bohdal and Fazl Barez were supported by the EPSRC Centre for Doctoral Training in Data Science (grant EP/L016427/1) and EPSRC Centre for Doctoral Training in Robotics and Autonomous Systems (EP/L016834/1) respectively, funded by the UK Engineering and Physical Sciences Research Council and the University of Edinburgh. Philip Torr was supported by Turing AI Fellowship EP/W002981/1.

**Conflict of Interest Statement:** N/A

**Informed Consent Statement:** N/A

**Acknowledgments:** We are grateful to Trond Undheim for providing highly useful and detailed suggestions for improving our paper as well as to Charlotte Siegmann and Shahar Avin for their feedback on our earlier draft. Their insights and comments have greatly improved the quality of this work.

**Author Contributions:** Ondrej Bohdal and Fazl Barez are the primary contributors. Timothy Hospedales and Philip Torr helped improve the ideas presented in the paper.

**Abstract:** Successful deployment of artificial intelligence (AI) in various settings has led to numerous positive outcomes for individuals and society. However, AI systems have also been shown to harm parts of the population due to biased predictions. AI fairness focuses on mitigating such biases to ensure AI decision making is not discriminatory towards certain groups. We take a closer look at AI fairness and analyze how lack of AI fairness can lead to deepening of biases over time and act as a social stressor. More specifically, we discuss how biased models can lead to more negative real-world outcomes for certain groups, which may then become more prevalent by deploying new AI models trained on increasingly biased data, resulting in a feedback loop. If the issues persist, they could be reinforced by interactions with other risks and have severe implications on society in the form of social unrest. We examine current strategies for improving AI fairness, assess their limitations in terms of real-world deployment, and explore potential paths forward to ensure we reap AI's benefits without causing society's collapse.

[1] Research Associate, School of Informatics, University of Edinburgh, Edinburgh, UK; ondrej.bohdal@ed.ac.uk.

[2] Professor, School of Informatics, University of Edinburgh, Edinburgh, UK and Head of Samsung AI Center, Cambridge, UK; t.hospedales@ed.ac.uk.

[3] Professor, Department of Engineering Science, University of Oxford, Oxford, UK; philip.torr@eng.ox.ac.uk.

[4] PhD Student, School of Informatics, University of Edinburgh, Edinburgh, UK and Visiting Student, Department of Engineering Science, University of Oxford, Oxford, UK; f.barez@ed.ac.uk.

* Correspondence: ondrej.bohdal@ed.ac.uk

## 1. Introduction

AI approaches offer excellent performance in many practically important problems (Caruana et al., 2015; Hinton et al., 2012; Andreu-Perez et al., 2018), but they can give biased and unfair predictions (Mehrabi et al., 2019; Hardt et al., 2016; Dwork et al., 2012). AI is increasingly often deployed to high-stakes applications (Tolan et al., 2019; Larrazabal et al., 2020; Seyyed-Kalantari et al., 2021), where unfair predictions can lead to substantial disadvantage or harm to parts of the population. For example, AI has been used for deciding who to select for interviews (Cohen et al., 2020; Bogen and Rieke, 2018), who should be given a mortgage (Martinez and Kirchner, 2021) or who is more likely to repeat crime after leaving prison (Tolan et al., 2019; Dressel and Farid, 2018). Unfair decisions in such key areas can have a significant impact on one's future.

We study the long-term social implications of unfair AI, from the perspective of continuous bias amplification stemming from new AI models trained on increasingly biased data. Current AI models have been shown to be biased, especially because of being trained on biased data (Mehrabi et al., 2019). At the same time, they have been shown to make more biased decisions than present in the training data, hence amplifying the biases (Lloyd, 2018; Hall et al., 2022).

Decisions of AI models influence the real world, and information about the real world is used for training new AI models. This means that more biased decisions will lead to more biased data for training new AI models, resulting in a more biased new generation of AI models. Additionally, parts of the population can experience bias from several sources, e.g. hiring and healthcare, and these combined can also put certain groups in increasingly large disadvantage over time. Overall this represents a feedback loop where new AI models give more and more biased decisions.

It is relatively common that AI biases are uncovered only after several years since the systems have been deployed (Dressel and Farid, 2018; Obermeyer et al., 2019; Martinez and Kirchner, 2021). Once a bias becomes recognized, action should be taken to rectify it. There are two main solutions: avoid using the biased AI model, or fix the AI model so that it is no longer biased. In many circumstances, it can be difficult to stop using the AI system as the system may already be deeply embedded and alternative solutions can be costly (Viechnicki and Eggers, 2017) or unviable, for example due to shortage of employees (Ford, 2021). On the other hand, improving the fairness of the AI model may take considerable time given that it is an open research problem (Mehrabi et al., 2019; Zong et al., 2023). Perhaps one of the best ways to mitigate bias is to curate bias-free data (Jordon et al., 2022), but this has its own challenges, especially because using real already collected data is significantly easier. Most suitable solutions to rectify biases depend on the specific circumstances, but in many cases rectifying AI bias can be challenging (Hao, 2019). As a result, the system may continue to be operating, perhaps with only minor changes. It is also important to note though that we should not completely avoid using AI as using it can bring numerous benefits and valuable insights – the key is to use AI responsibly.

Nevertheless, if parts of the population are systematically marginalized because of biased AI models, they can be under severe stress and they may try to resolve the situation by protesting against the deployment of such AI systems. If the institutions find it challenging to stop using biased AI, e.g. due to lack of employees or resources more broadly, disadvantaged groups may resort to escalating the situation using violence. In this sense lack of AI fairness can act as a social stressor that can lead to social unrest if the issues are prevalent and not addressed.

Deployment of insufficiently fair AI systems would likely be only one of several social stressors. Consequently, we study the interaction with other stressors, especially climate change that has increasingly significant impact on the society. The interaction among multiple social stressors can reinforce each other and result in more extensive social unrest. Over longer time horizons, this could destabilize the situations in various countries.

In addition to studying social implications, we also investigate what approaches are being developed to improve AI fairness. We give particular focus on real-world deployment of fair AI models and identify lack of fairness generalization across data distribution shifts can be a key challenge. We discuss approaches for robust fairness, but we also discuss approaches from areas related to out-of-distribution robustness, including domain generalization and adaptation. We conclude with a broader discussion that includes suggestions that can help mitigate risks and obtain more benefits from deploying AI.

**2. Definitions of Fairness**

Researchers have proposed a variety of ways to define fairness (Mehrabi et al., 2019), and some of the most common include equalized odds (Hardt et al., 2016), equal opportunity (Hardt et al., 2016) and demographic parity (Dwork et al., 2012). The unifying theme of these metrics is we want to ensure the same or similar probability of the given outcomes across all considered groups, which we illustrate in Figure 1. The concept of AI fairness hence captures the notion of ensuring AI decision making is not discriminatory toward any of the groups that interact with the AI system.

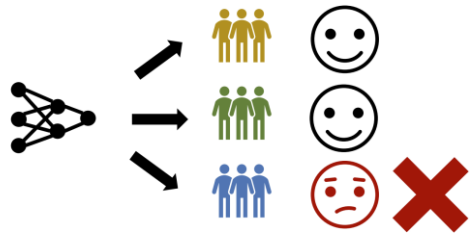

*Figure 1: We focus on the topic of fairness where we want to ensure that all groups receive unbiased and equal treatment so that no groups are harmed because of using AI.*

Equalized odds (Hardt et al., 2016) for predictor $\hat{Y}$, target $Y$ and protected attribute $A$ are defined for the binary case as:

$$Pr\{\hat{Y} = 1 | A = 0, Y = y\} = Pr\{\hat{Y} = 1 | A = 1, Y = y\}, \quad y \in \{0,1\}.$$

The definition means $\hat{Y}$ has equal true positive rate for demographics $A = 0$ and $A = 1$ if the outcome is $y = 1$, and equal false positive rates if the outcome is $y = 0$.

Equal opportunity (Hardt et al., 2016) is a relaxed alternative of equalized odds as it only requires non-discrimination within the advantageous outcome group:

$$Pr\{\hat{Y} = 1 | A = 0, Y = 1\} = Pr\{\hat{Y} = 1 | A = 1, Y = 1\}.$$

Compared to earlier demographic parity metric (Dwork et al., 2012), the benefit of equalized odds and equal opportunity is that they do not require independence from the protected attribute (Hardt et al., 2016). More broadly when deciding which metric to use, it is key to consider the suitability for the specific application (Zong et al., 2023).

In addition to specialized fairness metrics, we can monitor the worst-case performance alongside the average performance (Zhang et al., 2021; Bohdal et al., 2022b). More specifically we can measure the performance on the most challenging group (Zhang et al., 2021) or if the notion is less clear, we can use e.g. the most challenging 10% of the examples used for evaluation (Bohdal et al., 2022b). Such way of evaluation can also be used for settings where we want to ensure fairness when deploying AI systems across different scenarios. It is related to Max-Min fairness (Lahoti et al., 2020) where a model with smaller worst-case error is seen as fairer.

We further consider a stronger notion of fairness that is important when deploying models to the real-world: fairness that is robust under data distribution shifts. Data distribution shifts are common when deploying models to the real world. They can arise for example from applying the model to populations of different characteristics than were used during training, or also from processing the data differently (e.g. due to using different cameras or sensors). A real world example is training an AI model using medical images taken with a scanner of one brand, and then deploying the model to a hospital where a scanner of different brand is used. AI models should not discriminate against any of the subgroups when deployed to real-world "in-the-wild" scenarios. We illustrate robust fairness in Figure 2, and we will also consider this notion when discussing current solutions towards AI fairness.

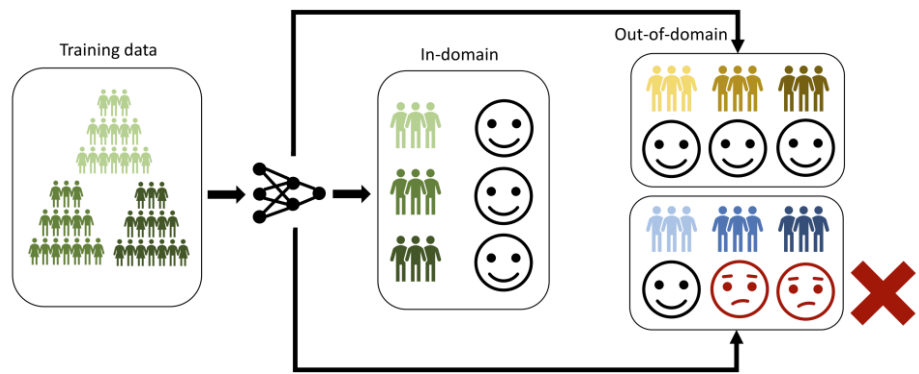

*Figure 2: When deploying AI models to the real world, it is crucial to ensure the models are robust and generalize fairness also to out-of-domain ("in-the-wild") situations.*

### 3. Social Implications of AI Fairness

We begin our analysis of social implications of AI fairness by introducing several high-stakes real-world examples where AI has already been used. We will then present a self-reinforcing feedback loop mechanism where biased AI systems lead to more biased outcomes, which then act as input data for further training of new AI systems. Over long periods of time this may lead to increasingly systemic social and economic marginalization of parts of the population. Such systemic marginalization could later become a substantial social stressor.

### 3.1 High-Stakes Real-World Applications

With the possibility to automate various time-consuming tasks and potentially improve upon imperfections of human decision-making, AI has been used for a number of high-stakes applications where fairness is important (Mehrabi et al., 2019). However, in many cases it has already been identified that the AI is unfair and causes harm to certain groups.

High-stakes real-world applications where fairness matters and has already been compromised include the following:

- Hiring for jobs: biased AI has been used in the context of hiring in multiple ways, including filtering of CVs (Cohen et al., 2020; Bogen and Rieke, 2018), evaluating video interviews (Kelly-Lyth, 2021) and delivering advertisements promoting jobs (Lambrecht and Tucker, 2019). Type of employment has a large impact on one's future, so it is key to ensure certain groups are not systematically disadvantaged (or given an advantage).
- Finance: AI can simplify the task of assessing if someone is likely to repay a loan or a mortgage, so such systems have already been deployed in practice. It has been shown that systems for making decisions about loans (Mukerjee et al., 2002) or mortgages (Martinez and Kirchner, 2021) can be significantly biased, for example making applicants of color 40 to 80% more likely to be denied mortgage application compared to white applicants (Martinez and Kirchner, 2021). If a group of certain characteristics is unable to get a mortgage and is forced to rent, it can have a large impact on their well-being, especially if it means they have to find new accommodation often.
- Public safety: unfair AI systems have been used in various public safety contexts, including sentencing decisions (Dressel and Farid, 2018; Tolan et al., 2019) and children welfare (Chouldechova et al., 2018). More specifically, AI has been used to predict the risk of recidivism as part of the COMPAS system (Dressel and Farid, 2018), and also as part of a tool applied to the particularly sensitive case of predicting juvenile recidivism (Tolan et al., 2019). Biased AI has also been used in the context of children welfare to perform screening of referrals for child protection (Chouldechova et al., 2018).
- Healthcare: biased AI systems have been deployed for multiple healthcare applications. For example, health-management systems (Obermeyer et al., 2019) have been shown to assign the same risk to black patients that are sicker than white patients. Biased AI has also led to underdiagnosis of under-served patient populations when applying AI to chest radiographs (Seyyed-Kalantari et al., 2021), and it also resulted in gender-biased computer-aided diagnosis (Larrazabal et al., 2020).

Additionally, it is not only the high-stakes situations where AI has the potential to discriminate and treat people unfairly. There are also situations where unfair AI can cause inconvenience. However, these can potentially act as a reminder that the person may have been treated unfairly by AI in some of the high-stakes situations, raising anger levels.

Face recognition is one of the key areas that exemplifies such scenarios and shows the need for fair and robust AI models. For instance, earlier facial processing systems from leading tech companies performed significantly worse on black women than on white men (Buolamwini and Gebru, 2018; Raji et al., 2020). Widespread use of such models could lead to frequent inconveniences, for example if face recognition technology were utilized for workplace access or for identifying criminals in public spaces.

The concept of fairness is also important in the context of generative language models such as GPT-4 (OpenAI, 2023), LaMDA (Thoppilan et al., 2022) and LLaMA (Touvron et al., 2023), because the generated content can influence people and have real-world impact. We want to ensure the generated content is not biased and does not include prejudices about any parts of the population. Further we want to ensure that safeguards in the models are robust across different languages and cultures so that we do not risk, for example, harming parts of the society in countries that speak lower-resource languages.

### 3.2 Self-Reinforcing Feedback Loop

Deployed AI models influence the society, and the outcomes form part of the training data for a new generation of AI models. If the initial models are biased, they produce biased outputs that will be used for training newer models. It has been shown that AI models amplify biases (Lloyd, 2018; Hall et al., 2022), which means new AI models would be biased even more due to training on increasingly biased data. We illustrate this in Figure 3, where we show the resulting feedback loop. Because many biases become evident only after wide deployment of the system, it is key to consider what long-term implications AI-amplified biases could have.

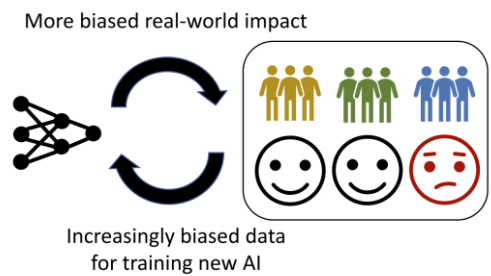

*Figure 3: Biased real-world outcomes lead to increasingly biased data for training new AI models, resulting in a self-enforcing feedback loop.*

Let us explain the bias amplification on examples. For example, if jobs hiring decision AI is based on past hiring decisions, then bias against subgroups in the past can reinforce to more bias in the future. If criminal sentencing AI is biased against a subgroup, that subgroup has longer prison sentences, which may make them harder to re-integrate with society after release. This may increase their likelihood of repeat crime recidivism, which will be data that increases the bias of the sentencing-decision AI the next time it is trained.

A whole ecosystem of AI models that happened to be biased against a particular subgroup could lead to persistently worse social and economic outcomes for that subgroup. More specifically, it could lead to worse jobs, worse access to finance, longer sentences for equivalent crimes, worse health due to worse medical treatment, worse educational outcomes if the education AI is biased. These biases then reinforce each other as e.g. worse health reinforces worse education and jobs. Over long periods of time that subgroup could become increasingly systemically socially and economically marginalized, which could become a substantial social stressor.

There is a risk of compounding of negative effects due to the feedback potential between the AI system decisions and real-world data, which affects the training data for the next round of AI training. The level of bias may become increasingly more difficult to tolerate and protests will arise. Moreover, as the technology becomes widespread, infrastructure will be built around it, so it will be challenging to remove it from use even as people protest against it. This would ultimately create tensions.

### 3.3 Fairness Risk

We present a toy model to estimate the fairness risk at time $t$ after deployment of the first AI systems. The formula models the compounding of biases over time (similar to compounding of interest rates):

$$\text{risk} \sim \beta^t,$$

where $\beta$ is the bias amplification rate of the AI models. When no bias amplification happens, $\beta = 1$, but because AI models have been shown to increase biases, typically $\beta > 1$. Over time the biases would amplify each other via repeated training of new AI systems on increasingly biased data. Such amplification of biases can lead to dissatisfaction with the AI systems and can act as a social stressor. Ideally we would like to have $\beta \leq 1$.

### 3.4 Challenges with Avoiding AI Automation

Budgets are typically tight (Tujula and Wolswijk, 2004), and ways to save resources are sought. AI offers ways to decrease costs (Viechnicki and Eggers, 2017; Berglind et al., 2022) and if there is a crisis, institutions may be more likely to use experimental approaches that have not been fully assessed. If the technology is likely to benefit most of the population, it may be difficult to argue against its deployment, especially if it is hard to measure potential harm it may cause. In these cases, it is important to recognize effects across different groups. One can try to mitigate them by investing more resources into making the technology fair and robust. It is also important to note that AI-based solutions may become deeply embedded in the software infrastructure and removing them once new significant biases are identified or if people start protesting, can be challenging.

In addition, automation using AI is not only about savings, it may also be inevitable due to shortages of employees to perform specific jobs (Ford, 2021). For example, shortages have been reported in areas as diverse as construction, social work, and transportation (Francis-Devine and Buchanan, 2023). Shortages may not always be resolved by increasing the budgets because some jobs require hard-to-obtain skills or cause significant amount of distress, so using AI would be desirable also because of non-monetary reasons. More generally AI can bring significant benefits and improve lives of people in various directions. As a result, it is important to find a suitable trade-off between AI use and regulation that tries to mitigate the negative impacts.

### 3.5 Interaction with Other Risks and Existential Implications

Unfair AI would be only one factor that would contribute to tension in the society. Another key driver of tension is likely to be climate change (Institute for Economics & Peace, 2020) and its implications (Nardulli et al., 2015), which include rising food prices or having to cover the costs of disaster responses. For example, large-scale drought in Syria is thought to have contributed to social stressors, which eventually led to an uprising in 2011 (Kelley et al., 2015). Syria is in civil war since 2011, already for more than a decade (Loft et al., 2022; Khen et al., 2020). The case of Syria shows that a combination of multiple stressors that lead to uprisings can result in a long-term civil war, which is a prime example of country's crisis. More broadly systemic unfairness, inequality, and marginalization of parts of the population has a record of leading to radicalization (van den Bos, 2020), violent uprisings and in some cases destabilization of societies. High-profile examples where these were likely to be a factor include the French revolution (Tocqueville, 1856), Indian independence movement (Chandra, 1989) and recently the Arab Spring (Haas, 2017).

A combination of multiple social stressors such as climate change and biased AI are likely to reinforce each other, which we illustrate in Figure 4. Persistent deployment of biased AI in high-stakes applications can lead to increasing levels of tension in the society and erode trust in the institutions. At a certain level it may be significant enough that in interaction with other social stressors such as climate change it escalates. It is crucial to try to mitigate the social stressors to avoid compounding effects.

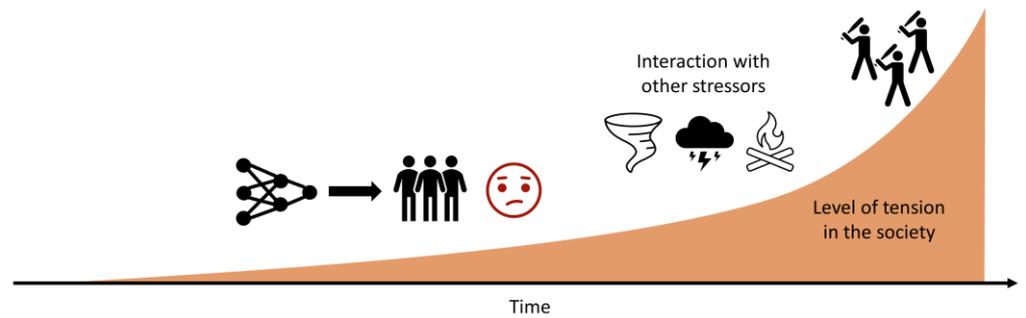

*Figure 4: Biased AI outcomes in interaction with other social stressors can lead to increasing levels of tension in the society and escalate if not mitigated.*

Biased AI, climate change and other social stressors have the potential to be commonplace across many nations, which brings the risk of widespread social unrest. Social unrest can turn into a civil war in severe cases (Kelley et al., 2015; Khen et al., 2020), which could potentially endanger the civilization if present in many countries. Overall this means that the risks from unfair AI can be large if it is deployed to too many critical applications without careful consideration.

**4. How to Improve AI Fairness?**

**4.1 Approaches for Fairness**

A large number of approaches for fairness has been proposed (Mehrabi et al., 2019; Zong et al., 2023), reflecting the importance of the field. Many of the recent methods try to alleviate bias as part of training the models, also known as in-processing (Mehrabi et al., 2019; Zong et al., 2023). Key in-processing families of bias mitigation methods include:

- Subgroup rebalancing (Chawla et al., 2002; Idrissi et al., 2022) that over-samples minority groups and down-samples majority groups,
- Domain independence (Wang et al., 2020; Royer and Lampert, 2015) that uses separate classifiers for different subgroups,
- Adversarial training (Madras et al., 2018; Zhao et al., 2019; Kim et al., 2019) that tries to train representations that make it difficult to identify different groups,
- Disentanglement (Tartaglione et al., 2021; Sarhan et al., 2020) methods that separate sensitive attributes and the useful attributes when constructing the representations.

Other families of methods for improving fairness also exist. Zong et al. (2023) have identified that domain generalization approaches (Sagawa et al., 2020; Cha et al., 2021; Foret et al., 2021) can be useful for improving fairness. Domain generalization methods try to learn representations that directly generalize to new out-of-domain situations without any adaptation, which relates to the goal of obtaining strong performance across different groups present in the population. Further, pre-processing methods (Khodadadian et al., 2021) try to remove bias from the dataset before training, for example by distorting the data (Khodadadian et al., 2021) as simply removing the sensitive attributes has been shown to be insufficient (Mehrabi et al., 2019). Post-processing methods (Pleiss et al., 2017) modify the predictions of an already trained model to improve fairness with respect to the sensitive attributes.

### 4.2 Fairness Under Data Distribution Shifts

When deploying AI models to the real-world, it is key to ensure the key properties of AI models hold also for real "in-the-wild" data. Such data are likely to come also from data distributions different from ones seen during training, so robustness against distribution shifts is crucial. For example, healthcare AI systems can be trained on data from selected prestigious US hospitals and deployed to hospitals of various quality across the US.

However, it has been shown that most existing fairness methods are only designed for in-domain settings and fail when data distribution changes (Schumann et al., 2019; Mishler and Dalmasso, 2022; Zong et al., 2023). Several approaches have been developed to tackle the challenge of fairness under distribution shift (Singh et al., 2021; Rezaei et al., 2020; Schumann et al., 2019), but these consider adaptation to a specific domain, with only (Pham et al., 2023) presenting an approach that generalizes across domains. If institutions only consider if an AI system is fair for in-domain data, such AI models may still lead to significant biases when deployed in the real-world and cause harm to parts of the population.

It has been identified (Zong et al., 2023) that domain generalization approaches (Sagawa et al., 2020; Cha et al., 2021; Foret et al., 2021) can offer competitive performance in terms of fairness, so improving fairness of domain generalization methods can be a good way forward. However, it has also been shown that domain generalization is a challenging problem on its own (Gulrajani and Lopez-Paz, 2021), with many approaches performing similarly as simple training across many domains (Vapnik, 1998) if following a fair evaluation protocol. As a result, domain adaptation approaches that adapt pre-trained models to local data distributions can be more successful in terms of maintaining fairness and strong performance. Source-free domain adaptation (Liang et al., 2020; Kundu et al., 2020; Ishii and Sugiyama, 2021; Yang et al., 2021) in particular can be practically valuable as it adapts a pre-trained model solely using unlabelled target domain data, without access to the source data. Efficient feed-forward approaches that perform adaptation without back-propagation (Bohdal et al., 2022b; Schneider et al., 2020; Bohdal et al., 2022a, 2023) can be especially useful on deployed devices.

### 4.3 Evaluation

Various benchmarks have been developed or repurposed for evaluating fairness. Key tabular fairness datasets include COMPAS (Dieterich et al., 2016), Adult Census (Ding et al., 2021; Kohavi and Becker, 1996) and Diabetes (Strack et al., 2014), some of which are also available within the popular Fairlearn library (Bird et al., 2020). Common computer vision fairness datasets include CIFAR-10S (Wang et al., 2020), CelebA (Liu et al., 2015) and IMDB face dataset (Rothe et al., 2015). Medical imaging datasets (Irvin et al., 2019; Johnson et al., 2019; Groh et al., 2021) have also been used extensively for evaluating fairness, with MEDFAIR (Zong et al., 2023) providing a suite of benchmarks to provide rigorous evaluation of fairness algorithms, including in-domain and out-of-domain scenarios.

Long term we believe it is key to develop new more extensive benchmarks that test both in-domain and out-of-domain scenarios, similar in scope to MEDFAIR (Zong et al., 2023) but covering various areas for which AI fairness is crucial. Because it has been observed that real-world datasets are often biased, synthetic datasets may be highly useful in the future. Synthetic data would enable us to design what unbiased outcomes look like and train models on them, improving fairness and robustness (Jordon et al., 2022). Once the model is trained with synthetic data, it can be fine-tuned using curated real-world data that do not need to be as plentiful.

### 4.4 Mechanistic Interpretability and Fairness

In the context of fairness, mechanistic interpretability (MI) is of significant importance. As a growing sub-field of interpretability, it aims to understand individual neurons within models and their larger circuits, playing an essential role in various applications (Olah et al., 2020b; Olah, 2022; Goh et al., 2021). The ultimate goal is to decompose a model into interpretable components, enabling more significant insights into model safety and bias detection (Hendrycks et al., 2021; Amodei et al., 2016; Vig et al., 2020). MI can be helpful in numerous fields such as autonomous vehicles (Barez et al., 2022) and Large Language Models (LLMs) (Miceli-Barone et al., 2023). However, fully achieving mechanistic interpretability in these domains remains a challenging task. One approach used in image models is feature visualization, where synthetic input images are employed to optimize the understanding of a target neuron (Olah et al., 2017). These techniques have significantly enhanced the interpretation of vision models, helping identify multimodal neurons that respond to abstract concepts (Goh et al., 2021) and cataloguing the behaviour of early neurons in Inception-v1 (Olah et al., 2020a). Examining individual neurons can allow for better and more human-readable visualizations (Foote et al., 2023), enabling testing of the role of scale, the addition of more data and parameters, and assessing the impact of removing undesired concepts (Hoelscher-Obermaier et al., 2023).

### 5. Discussion

### 5.1 Broader AI Safety

We have explored how insufficiently fair AI could contribute to the collapse of society if it is delegated too many critical systemic functions. Our approach is relatively unique in the space of existential-risk literature because we focus on discussing the danger of AI that has already become commonplace, rather than focusing on highly-intelligent AI having goals misaligned with human values. We believe both present risks and deserve significant attention. To provide a more complete view of AI safety, we briefly discuss also the other directions studying AI risks.

A variety of hazards have been associated with advanced AIs, including weaponization, proxy gaming, emergent goals, deception, and power-seeking behaviour (Hendrycks and Mazeika, 2022). It has been discussed how due to competitive pressures the most successful AIs would have by default undesirable traits, ones misaligned with human values (Hendrycks, 2023). Consequently, it can be expected that without suitable interventions, those hazards could become real when AI reaches sufficiently advanced level. A variety of high-level interventions have been proposed, including careful design of AI's intrinsic motivations, constraints on actions and cooperative institutions.

Another serious risk comes from the fact that even the current AI models have the scope for malicious use, with the ability to impact digital, physical, and also political security (Brundage et al., 2018). Current AI can be dual-use, for example with the potential to develop new chemical weapons (Urbina et al., 2022). Existing generative models also have the scope for introducing serious hazards, for example by being used for spreading misinformation on a large scale (Goldstein et al., 2023).

### 5.2 Recommendations

Considering the significant impact that insufficient AI fairness can have on the future of our society, there are various steps that can be taken to mitigate negative impact and make AI more beneficial. We identify three key areas that we discuss in more depth.

### 5.2.1 Develop the Science of Iterative Bias Amplification

Existing work has shown that AI models have the tendency to amplify biases (Lloyd, 2018; Hall et al., 2022). These biases influence the real-world, which introduces the feedback loop where the biases become stronger over time. To better analyze the impact of deploying imperfect AI models over longer time horizons, it would be useful to study this from various perspectives. More specifically, we suggest developing a science of iterative bias amplification that will help us understand how decisions made by current AI systems (which determine the training set of future AI systems) affect the evolution of AI bias and fairness in the long run. Agent-based modelling (Bonabeau, 2002) could be a useful tool for studying this area.

### 5.2.2 Develop Foundational Synthetic Datasets

One of the main reasons why AI models are biased is that they are trained on biased data that reflect biases present in the society (Mehrabi et al., 2019). A suitable solution could be to develop new foundational synthetic datasets that can be used for fair pre-training of AI models (Jordon et al., 2022). Constructing such large-scale datasets can be expensive, not only due to the size but also because various parties would need to be involved in the design process to ensure usefulness of the datasets. In a way the design of such datasets could benefit from consultations similar to the ones used for designing new policies. Consequently, we envision it would be primarily the governments and other large institutions that would cover the costs of these public-benefit datasets.

### 5.2.3 Policy Guidelines and Regulations

AI is likely to have increasingly large real-world impact, so it is crucial to ensure adequate policy guidelines and regulations regarding AI fairness are in place. This includes policy guidelines and regulations that have real-world impact on funding, algorithmic transparency, and which mandate human-in-the-loop for important applications. As the society evolves over time, it will be key to also require monitoring of deployed AI systems as they influence people in the real-world. The biases may not be present during initial evaluation, but can arise later after the system has been deployed. It is encouraging to see governments are introducing policies on fair and responsible use of AI (European Commission, 2020; Office of Science and Technology Policy, 2022; Department for Science, Innovation and Technology, 2023), but it will be key to ensure the policies are continuously adapted as the society evolves and new research is conducted.

### 6. Conclusion

In this paper we have investigated the long-term implications of unfair AI systems. We have identified that a feedback loop that leads to increasingly large biases can arise as biased AI models impact the population and new AI models are trained on such outcomes. Over longer time horizons, increasing levels of systemic unfairness can act as a social stressor and trigger protests, which can result in social unrest. We have discussed real-world limitations of existing AI systems designed to be fair and suggested steps that can be taken to improve the situation. Overall, we believe that thanks to the significant interest from both the ML community and institutions deploying AI systems, potential severe risks stemming from biased AI systems can be avoided, but carefulness and extensive further research will be key.

**References**

Amodei, D., Olah, C., Steinhardt, J., Christiano, P., Schulman, J., and Mané, D. (2016). Concrete problems in AI safety. ArXiv.

Andreu-Perez, J., Deligianni, F., Ravi, D., and Yang, G.-Z. (2018). Artificial intelligence and robotics. Technical report, UK-RAS Network.

Barez, F., Hasanbeig, H., and Abate, A. (2022). System III: Learning with domain knowledge for safety constraints. In NeurIPS ML Safety Workshop.

Berglind, N., Fadia, A., and Isherwood, T. (2022). The potential value of AI—and how governments could look to capture it. Technical report, McKinsey & Company.

Bird, S., Dudík, M., Edgar, R., Horn, B., Lutz, R., Milan, V., Sameki, M., Wallach, H., and Walker, K. (2020). Fairlearn: A toolkit for assessing and improving fairness in AI. Technical report, Microsoft.

Bogen, M. and Rieke, A. (2018). Help wanted: an examination of hiring algorithms, equity, and bias. Technical report, Upturn.

Bohdal, O., Li, D., and Hospedales, T. (2022a). Feed-forward source-free domain adaptation via class prototypes. In ECCV OOD-CV Workshop.

Bohdal, O., Li, D., and Hospedales, T. (2023). Label calibration for semantic segmentation under domain shift. In ICLR Workshop on Trustworthy ML.

Bohdal, O., Li, D., Hu, S. X., and Hospedales, T. (2022b). Feed-forward source-free latent domain adaptation via cross-attention. In ICML Pre-training Workshop.

Bonabeau, E. (2002). Agent-based modeling: Methods and techniques for simulating human systems. Proceedings of the National Academy of Sciences.

Brundage, M., Avin, S., Clark, J., Toner, H., Eckersley, P., Garfinkel, B., Dafoe, A., Scharre, P., Zeitzoff, T., Filar, B., Anderson, H., Roff, H., Allen, G. C., Steinhardt, J., Flynn, C., hÉigeartaigh, S. Ó., Beard, S., Belfield, H., Farquhar, S., Lyle, C., Crootof, R., Evans, O., Page, M., Bryson, J. J., Yampolskiy, R. V., and Amodei, D. (2018). The malicious use of artificial intelligence: forecasting, prevention, and mitigation. ArXiv.

Buolamwini, J. and Gebru, T. (2018). Gender shades: Intersectional accuracy disparities in commercial gender classification. In FAccT.

Caruana, R., Lou, Y., Gehrke, J., Koch, P., Sturm, M., and Elhadad, N. (2015). Intelligible models for healthcare. In KDD.

Cha, J., Chun, S., Lee, K., Cho, H.-C., Park, S., Lee, Y., and Park, S. (2021). SWAD: Domain generalization by seeking flat minima. In NeurIPS.

Chandra, B. (1989). India's struggle for independence, 1857-1947. Penguin Books, New Delhi, India.
Chawla, N. V., Bowyer, K. W., Hall, L. O., and Kegelmeyer, W. P. (2002). SMOTE: Synthetic minority over-sampling technique. Journal of Artificial Intelligence Research.

Chouldechova, A., Benavides-Prado, D., Fialko, O., and Vaithianathan, R. (2018). A case study of algorithm-assisted decision making in child maltreatment hotline screening decisions. In FAccT.

Cohen, L., Lipton, Z. C., and Mansour, Y. (2020). Efficient candidate screening under multiple tests and implications for fairness. In Symposium on Foundations of Responsible Computing.

Department for Science, Innovation and Technology (2023). A pro-innovation approach to AI regulation. Technical report, UK Goverment.

Dieterich, W., Mendoza, C., and Brennan, T. (2016). COMPAS risk scales: Demonstrating accuracy equity and predictive parity. Technical report, Northpointe.

Ding, F., Hardt, M., Miller, J., and Schmidt, L. (2021). Retiring adult: New datasets for fair machine learning. In NeurIPS.

Dressel, J. and Farid, H. (2018). The accuracy, fairness, and limits of predicting recidivism. Science Advances.

Dwork, C., Hardt, M., Pitassi, T., Reingold, O., and Zemel, R. (2012). Fairness through awareness. In ITCS.

European Commission (2020). On Artificial Intelligence - A European approach to excellence and trust. Technical report, European Union.

Foote, A., Nanda, N., Kran, E., Konstas, I., Cohen, S., and Barez, F. (2023). Neuron to graph: Interpreting language model neurons at scale. In ICLR RTML Workshop.

Ford, M. (2021). Robots: stealing our jobs or solving labour shortages? Guardian.

Foret, P., Kleiner, A., Mobahi, H., and Neyshabur, B. (2021). Sharpness-aware minimization for efficiently improving generalization. In ICLR.

Francis-Devine, B. and Buchanan, I. (2023). Skills and labour shortages. Technical report, House of Commons Library.

Goh, G., Cammarata, N., Voss, C., Carter, S., Petrov, M., Schubert, L., Radford, A., and Olah, C. (2021). Multimodal neurons in artificial neural networks. Distill.

Goldstein, J. A., Sastry, G., Musser, M., DiResta, R., Gentzel, M., and Sedova, K. (2023). Generative language models and automated influence operations: Emerging threats and potential mitigations. ArXiv.

Groh, M., Harris, C., Soenksen, L., Lau, F., Han, R., Kim, A., Koochek, A., and Badri, O. (2021). Evaluating deep neural networks trained on clinical images in dermatology with the fitzpatrick 17k dataset. In CVPR.

Gulrajani, I. and Lopez-Paz, D. (2021). In search of lost domain generalization. In ICLR.

Haas, M. L. (2017). The Arab Spring: The hope and reality of the uprisings. Routledge, New York, USA.

Hall, M., van der Maaten, L., Gustafson, L., and Adcock, A. B. (2022). A systematic study of bias amplification. ArXiv.

Hao, K. (2019). This is how AI bias really happens—and why it's so hard to fix. MIT Technology Review.

Hardt, M., , Price, E., and Srebro, N. (2016). Equality of opportunity in supervised learning. In NIPS.

Hendrycks, D. (2023). Natural selection favors AIs over humans. ArXiv.

Hendrycks, D., Carlini, N., Schulman, J., and Steinhardt, J. (2021). Unsolved problems in ML safety. ArXiv.

Hendrycks, D. and Mazeika, M. (2022). X-risk analysis for AI research. ArXiv.

Hinton, G., Deng, L., Yu, D., Dahl, G., Mohamed, A.-r., Jaitly, N., Senior, A., Vanhoucke, V., Nguyen, P., Sainath, T., and Kingsbury, B. (2012). Deep neural networks for acoustic modeling in speech recognition: the shared views of four research groups. IEEE Signal Processing Magazine.

Hoelscher-Obermaier, J., Persson, J., Kran, E., Konstas, I., and Barez, F. (2023). Detecting edit failures in large language models: An improved specificity benchmark. In ACL Findings.

Idrissi, B. Y., Arjovsky, M., Pezeshki, M., and Lopez-Paz, D. (2022). Simple data balancing achieves competitive worst-group-accuracy. In Conference on Causal Learning and Reasoning.

Institute for Economics & Peace (2020). Ecological threat register. Technical report.

Irvin, J. A., Rajpurkar, P., Ko, M., Yu, Y., Ciurea-Ilcus, S., Chute, C., Marklund, H., Haghgoo, B., Ball, R. L., Shpanskaya, K. S., Seekins, J., Mong, D. A., Halabi, S. S., Sandberg, J. K., Jones, R., Larson, D. B., Langlotz, C., Patel, B. N., Lungren, M. P., and Ng, A. (2019). Chexpert: A large chest radiograph dataset with uncertainty labels and expert comparison. In AAAI.

Ishii, M. and Sugiyama, M. (2021). Source-free domain adaptation via distributional alignment by matching batch normalization statistics. ArXiv.

Johnson, A. E. W., Pollard, T. J., Berkowitz, S. J., Greenbaum, N. R., Lungren, M. P., ying Deng, C., Mark, R. G., and Horng, S. (2019). MIMIC-CXR: A large publicly available database of labeled chest radiographs. Scientific Data.

Jordon, J., Szpruch, L., Houssiau, F., Bottarelli, M., Cherubin, G., Maple, C., Cohen, S. N., and Weller, A. (2022). Synthetic data - what, why and how? ArXiv.

Kelley, C. P., Mohtadi, S., Cane, M. A., Seager, R., and Kushnir, Y. (2015). Climate change in the fertile crescent and implications of the recent Syrian drought. Proceedings of the National Academy of Sciences.

Kelly-Lyth, A. (2021). Challenging biased hiring algorithms. Oxford Journal of Legal Studies.

Khen, H. M.-E., Boms, N. T., and Ashraph, S. (2020). Introduction: An overview of stakeholders and interests. In The Syrian war: Between justice and political reality, pages 1–8. Cambridge University Press.

Khodadadian, S., Ghassami, A., and Kiyavash, N. (2021). Impact of data processing on fairness in supervised learning. In International Symposium on Information Theory (ISIT).

Kim, B., Kim, H., Kim, K., Kim, S., and Kim, J. (2019). Learning not to learn: Training deep neural networks with biased data. In CVPR.

Kohavi, R. and Becker, B. (1996). Adult data set. UCI Machine Learning Repository.

Kundu, J. N., Venkat, N., M, R., and Babu, R. V. (2020). Universal source-free domain adaptation. In CVPR.

Lahoti, P., Beutel, A., Chen, J., Lee, K., Prost, F., Thain, N., Wang, X., and Chi, E. H. (2020). Fairness without demographics through adversarially reweighted learning. In NIPS.

Lambrecht, A. and Tucker, C. (2019). Algorithmic bias? An empirical study of apparent gender-based discrimination in the display of STEM career ads. Management Science.

Larrazabal, A. J., Nieto, N., Peterson, V., Milone, D. H., and Ferrante, E. (2020). Gender imbalance in medical imaging datasets produces biased classifiers for computer-aided diagnosis. Proceedings of the National Academy of Sciences.

Liang, J., Hu, D., and Feng, J. (2020). Do we really need to access the source data? Source hypothesis transfer for unsupervised domain adaptation. In ICML.

Liu, Z., Luo, P., Wang, X., and Tang, X. (2015). Deep learning face attributes in the wild. In ICCV.

Lloyd, K. (2018). Bias amplification in artificial intelligence systems. ArXiv.

Loft, P., Sturge, G., and Kirk-Wade, E. (2022). The Syrian civil war: Timeline and statistics. Technical report, House of Commons Library.

Madras, D., Creager, E., Pitassi, T., and Zemel, R. (2018). Learning adversarially fair and transferable representations. In ICML.

Martinez, E. and Kirchner, L. (2021). The secret bias hidden in mortgage-approval algorithms. The Markup.

Mehrabi, N., Morstatter, F., Saxena, N. A., Lerman, K., and Galstyan, A. G. (2019). A survey on bias and fairness in machine learning. ACM Computing Surveys (CSUR).

Miceli-Barone, A. V., Barez, F., Konstas, I., and Cohen, S. B. (2023). The larger they are, the harder they fail: Language models do not recognize identifier swaps in python. In ACL.

Mishler, A. and Dalmasso, N. (2022). Fair when trained, unfair when deployed: Observable fairness measures are unstable in performative prediction settings. In NeurIPS Algorithmic Fairness through the Lens of Causality and Privacy Workshop.

Mukerjee, A., Biswas, R., Deb, K., and Mathur, A. P. (2002). Multi-objective evolutionary algorithms for the risk-return trade-off in bank loan management. International Transactions in Operational Research.

Nardulli, P. F., Peyton, B., and Bajjalieh, J. (2015). Climate change and civil unrest: The impact of rapid-onset disasters. The Journal of Conflict Resolution.

Obermeyer, Z., Powers, B., Vogeli, C., and Mullainathan, S. (2019). Dissecting racial bias in an algorithm used to manage the health of populations. Science.

Office of Science and Technology Policy (2022). Blueprint for an AI Bill of Rights: Making automated systems work for the American people. Technical report, The White House.

Olah, C. (2022). Mechanistic interpretability, variables, and the importance of interpretable bases. Transformer Circuits Thread.

Olah, C., Cammarata, N., Schubert, L., Goh, G., Petrov, M., and Carter, S. (2020a). An overview of early vision in inceptionv1. Distill.

Olah, C., Cammarata, N., Schubert, L., Goh, G., Petrov, M., and Carter, S. (2020b). Zoom in: An introduction to circuits. Distill.

Olah, C., Mordvintsev, A., and Schubert, L. (2017). Feature visualization. Distill.

OpenAI (2023). GPT-4 technical report. ArXiv.

Pham, T.-H., Zhang, X., and Zhang, P. (2023). Fairness and accuracy under domain generalization. In ICLR.

Pleiss, G., Raghavan, M., Wu, F., Kleinberg, J., and Weinberger, K. Q. (2017). On fairness and calibration. In NIPS.

Raji, I. D., Gebru, T., Mitchell, M., Buolamwini, J., Lee, J., and Denton, E. (2020). Saving face: Investigating the ethical concerns of facial recognition auditing. In AIES.

Rezaei, A., Liu, A., Memarrast, O., and Ziebart, B. D. (2020). Robust fairness under covariate shift. In AAAI.

Rothe, R., Timofte, R., and Van Gool, L. (2015). DEX: Deep expectation of apparent age from a single image. In ICCV Workshops.

Royer, A. and Lampert, C. H. (2015). Classifier adaptation at prediction time. In CVPR.

Sagawa, S., Koh, P. W., Hashimoto, T. B., and Liang, P. (2020). Distributionally robust neural networks. In ICLR.

Sarhan, M. H., Navab, N., Eslami, A., and Albarqouni, S. (2020). Fairness by learning orthogonal disentangled representations. In ECCV.

Schneider, S., Rusak, E., Eck, L., Bringmann, O., Brendel, W., and Bethge, M. (2020). Improving robustness against common corruptions by covariate shift adaptation. In NeurIPS.

Schumann, C., Wang, X., Beutel, A., Chen, J., Qian, H., and Chi, E. H. (2019). Transfer of machine learning fairness across domains. In NeurIPS AI for Social Good Workshop.

Seyyed-Kalantari, L., Zhang, H., McDermott, M. B. A., Chen, I. Y., and Ghassemi, M. (2021). Underdiagnosis bias of artificial intelligence algorithms applied to chest radiographs in underserved patient populations. Nature medicine.

Singh, H., Singh, R., Mhasawade, V., and Chunara, R. (2021). Fairness violations and mitigation under covariate shift. In FAccT.

Strack, B., DeShazo, J. P., Gennings, C., Olmo, J. L., Ventura, S., Cios, K. J., and Clore, J. N. (2014). Impact of HbA1c measurement on hospital readmission rates: Analysis of 70,000 clinical database patient records. BioMed Research International.

Tartaglione, E., Barbano, C. A., and Grangetto, M. (2021). EnD: Entangling and disentangling deep representations for bias correction. In CVPR.

Thoppilan, R., De Freitas, D., Hall, J., Shazeer, N., Kulshreshtha, A., Cheng, H.-T., Jin, A., Bos, T., Baker, L., Du, Y., Li, Y., Lee, H., Zheng, H. S., Ghafouri, A., Menegali, M., Huang, Y., Krikun, M., Lepikhin, D., Qin, J., Chen, D., Xu, Y., Chen, Z., Roberts, A., Bosma, M., Zhao, V., Zhou, Y., Chang, C.-C., Krivokon, I., Rusch, W., Pickett, M., Srinivasan, P., Man, L., Meier-Hellstern, K., Morris, M. R., Doshi, T., Santos, R. D., Duke, T., Soraker, J., Zevenbergen, B., Prabhakaran, V., Diaz, M., Hutchinson, B., Olson, K., Molina, A., Hoffman-John, E., Lee, J., Aroyo, L., Rajakumar, R., Butryna, A., Lamm, M., Kuzmina, V., Fenton, J., Cohen, A., Bernstein, R., Kurzweil, R., Aguera-Arcas, B., Cui, C., Croak, M., Chi, E., and Le, Q. (2022). LaMDA: Language models for dialog applications. ArXiv.

Tocqueville, A. (1856). The old regime and the revolution. Harper and Brothers, New York, USA.

Tolan, S., Miron, M., Gómez, E., and Castillo, C. (2019). Why machine learning may lead to unfairness: Evidence from risk assessment for juvenile justice in Catalonia. In International Conference on Artificial Intelligence and Law.

Touvron, H., Lavril, T., Izacard, G., Martinet, X., Lachaux, M.-A., Lacroix, T., Rozière, B., Goyal, N., Hambro, E., Azhar, F., Rodriguez, A., Joulin, A., Grave, E., and Lample, G. (2023). LLaMA: Open and efficient foundation language models. ArXiv.

Tujula, M. and Wolswijk, G. (2004). What determines fiscal balances? An empirical investigation in determinants of changes in OECD budget balances. SSRN Electronic Journal.

Urbina, F., Lentzos, F., Invernizzi, C., and Ekins, S. (2022). Dual use of artificial-intelligence-powered drug discovery. Nature Machine Intelligence.

van den Bos, K. (2020). Unfairness and radicalization. Annual Review of Psychology.

Vapnik, V. (1998). Statistical learning theory.

Viechnicki, P. and Eggers, W. D. (2017). How much time and money can AI save government? Technical report, Deloitte.

Vig, J., Gehrmann, S., Belinkov, Y., Qian, S., Nevo, D., Singer, Y., and Shieber, S. (2020). Investigating gender bias in language models using causal mediation analysis. In NeurIPS.

Wang, Z., Qinami, K., Karakozis, I., Genova, K., Nair, P., Hata, K., and Russakovsky, O. (2020). Towards fairness in visual recognition: Effective strategies for bias mitigation. In CVPR.

Yang, S., Wang, Y., van de Weijer, J., Herranz, L., and Jui, S. (2021). Exploiting the intrinsic neighborhood structure for source-free domain adaptation. In NeurIPS.

Zhang, M., Marklund, H., Dhawan, N., Gupta, A., Levine, S., and Finn, C. (2021). Adaptive risk minimization: learning to adapt to domain shift. In NeurIPS.

Zhao, H., Coston, A., Adel, T., and Gordon, G. J. (2019). Conditional learning of fair representations. In ICLR.

Zong, Y., Yang, Y., and Hospedales, T. (2023). MEDFAIR: Benchmarking fairness for medical imaging. In ICLR.